\documentstyle[epsfig,epsf,preprint,aps,eqsecnum]{revtex}

\newcommand{\lsim}{\mathrel{\lower4pt\hbox{$\sim$}}
\hskip-12.5pt\raise1.6pt\hbox{$<$}\;}
\newcommand{\gsim}{\mathrel{\lower4pt\hbox{$\sim$}}
\hskip-12.5pt\raise1.6pt\hbox{$>$}\;}

\newcommand{\bea}{\begin{eqnarray}}
\newcommand{\eea}{\end{eqnarray}}

\newcommand{\be}{\begin{equation}}
\newcommand{\ee}{\end{equation}}

\tightenlines

\begin{document}

\draft
\preprint{ULB-TH/99-17}

\title{QCD at $\theta \sim \pi$ reexamined:\\ Domain walls
and Spontaneous CP violation}

\author{Michel H.G. Tytgat\\ }

\address{Service de Physique Th\'eorique, CP225\\
Universit\'e Libre de Bruxelles\\
Bld du Triomphe, 1050 Bruxelles, Belgium}

\maketitle

\begin{abstract}
We consider  QCD  at $\theta \sim \pi$ with two, one and zero light flavours $N_f$,
using the Di Vecchia-Veneziano-Witten effective lagrangian.
For $N_f=2$, 
we show that CP is spontaneously broken at $\theta =\pi$  for finite quark mass splittings, $z=m_d/m_u \not= 1$.
In the $z-\theta$ plane, there is a line of first order 
transitions at $\theta=\pi$  with two critical endpoints,  $z_1^\ast < z < z_2^\ast$.
 We compute the tension of the domain walls that relate the two CP violating vacua. 
For  $m_u=m_d$,  the tension of the family of equivalent domain walls agrees with the expression 
derived by Smilga from chiral perturbation theory at next-to-leading order.
For $z_1^\ast < z < z^\ast_2$, $z\not= 1$,  there is only one domain wall and  a
wall-some sphaleron at $\theta = \pi$. 
At the critical points, $z = z_{1,2}^\ast$, the domain wall fades away, CP is restored
 and the transition becomes of second order.
For $N_f=1$, CP is spontaneously broken  
only if the number of colours $N_c$ is large and/or if the quark is sufficiently heavy. Taking the heavy quark limit 
($\sim N_f= 0$) provides a simple derivation
 of  the multibranch $\theta$ dependence of
the vacuum energy of large $N_c$  pure Yang-Mills theory. In the large $N_c$ limit, there are many quasi-stable vacua with decay rate $\Gamma \sim \exp(-N_c^4)$.

\end{abstract}

\newpage

\section{Introduction}

In the limit of $N_f$ massless quarks, QCD has 
a global $SU(N_f)\times SU(N_f)$ chiral symmetry. In vacuum, this symmetry is  spontaneously broken
to the diagonal $SU(N_f)$, with $N_f^2 -1$ massless Goldstone bosons. 
Introducing quark masses lifts the
degeneracy of the vacuum and gives a mass to the Goldstone bosons. Beside the quark
masses, there is  another parameter in QCD, known as Theta  ($\theta$). In Nature, 
$\theta \sim 0$ modulo $2\pi$. A non-zero value of
$\theta$ would introduces explicit CP violation in the strong interactions, with 
decay processes like $\eta^\prime \rightarrow \pi \pi$, or large contributions to the electric 
dipole moment of the 
neutron,  phenomena which are not observed~\cite{Crewther:1979pi}. Although large 
values of $\theta$ are excluded
experimentally, it is fruitful to investigate how strong interactions change for
$\theta\not=0$.  A beautiful example  is the  Veneziano-Witten formula which relates, in the limit
of large number of colours $N_c$,   the mass of the $\eta^\prime$ meson to the second derivative
with respect to $\theta$ of the vacuum energy of  pure Yang-Mills theory~\cite{Veneziano:1979ec,Witten:1979vv}. 

We will focus here on the fascinating, but unfortunately academic possibility of  
spontaneous CP violation at
$\theta = \pi$, also known as Dashen's phenonemon~\cite{Dashen:1971et}. 
 Strong interactions are {\em a priori} invariant under CP at   $\theta = \pi$ . This is because  
$\theta  = \pi 
\rightarrow -\,\pi$ under a CP transformation  but physics should be 
unchanged for $\theta \rightarrow \theta
+ 2 \pi$, so that  $\pi \equiv -\pi$. However, as shown by 
Dashen  some time before the advent of QCD,  CP can  be 
spontaneously broken at $\theta = \pi$, with the appearance of 
two  CP violating degenerate vacua 
separated by a
potential energy barrier ~\cite{Dashen:1971et}. According to the Vafa-Witten theorem, this possibility is
excluded at $\theta =0$~\cite{Vafa:1984xg,Azcoiti:1999rq}.
In the vicinity of $\theta =\pi$, one of the vacua has  lower energy and,  as  $\theta$ varies, there is  a first order
transition at $\theta = \pi$. 
This phenomenon and related issues has been
investigated  by Di Vecchia and 
Veneziano~\cite{DiVecchia:1980ve} and   Witten~\cite{Witten:1980sp} in the large  $N_c$ limit, 
and more recently by Creutz~\cite{Creutz:1995wf}, Evans {\em et al}~\cite{Evans:1997eq} and
Smilga~\cite{Smilga:1998dh}.

The realistic case of three light flavours has been the most discussed. 
Di Vecchia and Veneziano~\cite{DiVecchia:1980ve} and Witten~\cite{Witten:1980sp} have shown 
that there are two degenerate vacua  
at $\theta = \pi$, provided the following constraint is satisfied
\begin{equation} 
\label{witten}
{m_u m_d\over m_s} > \vert m_d - m_u\vert.
\label{ineqone}
\end{equation}
To our knowledge, the equivalent of~(\ref{ineqone}) for two light flavours 
has never been 
published, presumably because this question is {\em a priori} academic, as the 
inequality~(\ref{ineqone}) is not 
satisfied for realistic quark 
masses. (In Nature, $m_u\approx 4 MeV$, $m_d\approx 7 MeV$ and $m_s \approx 150 MeV$.)
The two flavour case is however quite interesting. On one hand, taking the limit $m_s\rightarrow
\infty$ in~(\ref{witten}) seems to imply that CP can only be broken for
$m_u=m_d$, which is in agreement with
the result found by Di Vecchia and Veneziano~\cite{DiVecchia:1980ve}. On the other hand Creutz~\cite{Creutz:1995wf} and
Evans {\em et al}~\cite{Evans:1997eq} found evidences of CP violation also for finite mass
splittings,  $z\equiv m_d/m_u \not=1$. 
The latter possibility is more natural. In the $z-\theta$ plane, we would expect a line of first order phase
 transitions at $\theta=\pi$,
terminated with critical endpoints $z=z^\ast$, where the phase transition becomes of second order.  
 A related issue, recently addressed by Smilga~\cite{Smilga:1998dh}, is that to leading order in chiral
perturbation theory ($\chi PT$) the potential term in 
\begin{equation}
\label{cpt}
{\cal L} = {f_\pi^2\over  4} \, \mbox{\rm tr}( \partial_\mu U^\dagger 
\partial^\mu U ) + \Sigma \, \mbox{\rm Re}\left[
\mbox{\rm tr} ({\cal M}\,e^{i \theta/2}\,U^\dagger)\right],
\end{equation}
where $U$ is an $SU(2)$ unitary matrix and with 
\begin{equation}
\label{mass}
    {\cal M} =  \left(
\begin{array}{cc}
m_u & 0 \\
0 & m_d \\
\end{array}\right ),
\end{equation}
vanishes at $\theta = \pi$ for $m_u=m_d$. This would imply that
pion excitations  are massless at $\theta = \pi$, with a second order phase
transition while it is expected to 
be of first order. 
Within $\chi PT$, the paradox is resolved  by taking into account next-to-leading order corrections in the 
quark mass  ${\cal O}(m^2)$, which lift the vacuum degeneracy at $\theta = \pi$~\cite{Smilga:1998dh}. 
However,  it is not manifest how precisely 
this is related to the phenomenon of spontaneous CP violation.

\bigskip

Our contribution  will be to draw a self-consistent picture of Dashen's phenomenon at $\theta=\pi$ for the case of 
two, one and zero light quark flavours. 
Incidentally, most (but
not all) of the results we will discuss  can be found 
scattered in the litterature cited above, either
explicitely or implicitely. 
We will work within the framework of the 
 large $N_c$ Di Vecchia-Veneziano-Witten effective lagrangian~\cite{DiVecchia:1980ve,Witten:1980sp,Rosenzweig:1980ay}.

For $N_f=1$, CP is spontaneously broken at $\theta =\pi$ only in the very large  $N_c$ limit or, alternatively, if the 
 quark  is sufficiently heavy. Increasing further the quark mass 
 provides a simple derivation, within field theory, of the
peculiar $\theta$ dependence of pure Yang-Mills theory at large $N_c$.
Recently, Witten~\cite{Witten:1998uk} 
has used the correspondence between large $N_c$ Yang-Mills theories and string theories on some particular 
compactified spacetimes to {\em derive} the qualitative form of the vacuum energy~\cite{Maldacena:1997re}. 
(See also~\cite{Shifman:1998if,Oz:1998fx,Gabadadze:1999na}.) 
Turning to  $N_f=2$, we will show that 
  CP is spontaneously broken 
with a first order transition at $\theta=\pi$ for a finite range of quark mass splittings,
$z_1^\ast < m_d/m_u < z_2^\ast$, and will  determine the critical values $z_{1,2}^\ast$ in the limit
$m_\pi^2 \ll m_{\eta^\prime}^2$. 
We will compute the tension of the domain walls relating the two CP violating vacua and, in the degenerate limit $z=1$, will 
 recover the result derived  by Smilga from chiral perturbation theory at next-to-leading order.
At the critical 
points $z = z_{1,2}^\ast$, the degenerate vacua merge and the domain wall disappears,  CP is restored 
and the phase transition becomes 
of second order. Chiral perturbation theory at leading order simply corresponds 
to the particular limit in which $z_1^\ast = z_2^\ast = 1$.

\section{Domain walls and spontaneous CP violation at $\theta = \pi$}
\label{secone}

In the large $N_c$ limit, the effects of the $U(1)_A$ anomaly 
fade away and the $\eta^\prime$ meson becomes light.
In particular, at infinite $N_c$ and in the chiral limit, the $\eta^\prime$ is massless and
there are $N_f^2$ Goldstone bosons. The phenomenological lagrangian that  incorporate both  quark mass
and   leading large $N_c$ effects 
is~\cite{DiVecchia:1980ve,Witten:1980sp,Rosenzweig:1980ay}, 
\begin{equation}
\label{dvw}
{\cal L} = {f_\pi^2\over  4} \, \mbox{\rm tr}( \partial_\mu U^\dagger 
\partial^\mu U ) + \Sigma \, \mbox{\rm Re}\left[
\mbox{\rm tr} ({\cal M}\,U^\dagger)\right] -{\tau \over 2} (\theta + i\, \log \det U)^2
\end{equation}
where, specializing to two light flavours,  ${\cal M}$ is the diagonal quark  mass
matrix~(\ref{mass}), $\Sigma = \vert \langle  \bar q q\rangle\vert$ and $\tau$ is the topological
susceptibility of pure  Yang-Mills theory. (In the real world, $\tau \sim (200 MeV)^4$.)
Because the mass matrix is diagonal, we can write 
the vacuum expectation value of the $U(2)$ matrix $U$  in the form
\begin{equation}
\label{uu}
U = \left(
\begin{array}{cc}
e^{i \phi_u} & 0 \\
0 & e^{i \phi_d} \\
\end{array}\right)
\end{equation}
If $m_u=m_d$, the choice~(\ref{uu})  for the vacuum expectation value of the unitary matrix $U$
is one among a whole manifold (with topology $SU(2)/U(1) \sim S^2$) of 
equivalent ones. This degeneracy is lifted for any $m_u\not= m_d$. 
We further decompose the phases $\phi_{u,d}$ as 
\begin{eqnarray*}
\phi_u &=& \phi + \alpha\nonumber\\
\phi_d &=& \phi - \alpha
\end{eqnarray*}
so that $\log\det U = 2\,i\, \phi$.
The phase $\phi$ is then related to the vacuum expectation value of the $\eta^\prime$ field 
\begin{equation}
\phi =  \langle\eta^\prime\rangle/f_\pi
\end{equation}
while $\alpha$ is the vacuum expectation value of the $\pi_0$ field. 
Also, in the chiral limit $m_u=m_d=0$, 
\begin{equation}
\label{vw}
f_\pi^2 \,m_{\eta^\prime}^2 = 4 \tau
\end{equation}
which is the celebrated Veneziano-Witten relation for two
flavours~\cite{Veneziano:1979ec,Witten:1979vv}.

In this basis, the potential energy term of~(\ref{dvw}) reads 
\begin{eqnarray}
\label{energy}
E(\theta) &=& - \Sigma m_u \cos (\phi + \alpha) - \Sigma m_d \cos(\phi - \alpha) + {\tau\over 2}
(\theta - 2\phi)^2\nonumber \\
&=& -(m_u + m_d) \Sigma \cos\phi\cos\alpha + (m_u - m_d)\Sigma \sin\phi\sin\alpha + {\tau\over2}(\theta -2 \phi)^2
\end{eqnarray}
Minimizing $E(\theta)$ with respect to $\phi$ and $\alpha$ gives the two equations
\begin{eqnarray}
\label{setone}
(m_u+m_d)\cos\phi\sin\alpha + (m_u-m_d)\sin\phi\cos\alpha & = &  0\\
\label{settwo}
(m_u+m_d)\sin\phi\cos\alpha + (m_u-m_d)\cos\phi\sin\alpha & = & {2 \tau\over\Sigma}\, (\theta
- 2\phi) 
\end{eqnarray}
%
For generic quark masses and $\theta\not= 0,\pi$,  the 
solutions of~(\ref{setone}--\ref{settwo}) are CP violating.

\subsection{General remarks}

If $m_d$ goes to zero, (\ref{setone}) gives $\phi + \alpha = 0$ modulo $\pi$, 
while~(\ref{settwo}) imposes $\phi=\theta/2$. (If $m_u \rightarrow 0$ instead, $\phi - \alpha=0$. In
the sequel,  we will keep $m_u$ fixed and vary $m_d$.) In essence, 
$\theta$ has been absorbed in the redefinition of the phase $\phi_d$ 
(which is unconstrained by the potential 
if $m_d=0$), and, as expected, there is no CP violation.
Another way to phrase this is that  
\begin{equation}
\bar \theta = (\theta - 2 \phi)\,\, \mbox{modulo}\,\, \pi
\end{equation}
give the measure of  $CP$ violation in strong interactions.\footnote{To see this, let 
 $U_0$ be the vacuum
expectation of the $U$ matrix,  and define $U = U_0 V$, $\bar {\cal M} = {\cal M} 
U_0 = {\cal A} + i {\cal B}$. 
As shown by Witten~\cite{Witten:1980sp}, vacuum stability requires 
\begin{equation}
\label{cool}
\Sigma \,{\cal B} =  2\tau \, \bar \theta \,\mbox{\bf 1}_2.
\end{equation}
With this decomposition and using~(\ref{cool}), the CP violating part of the 
potential~(\ref{dvw}) is
\begin{equation}
E(\theta)_{CP} = -  \,i\,\tau\,\bar\theta\left[ 2 \mbox{\rm tr}
(\mbox{\rm Im}V) -  \,\log\det V\right] 
\end{equation}  
which vanishes if $\bar\theta = 0$.}

In the opposite limit of a decoupling heavy quark, $m_d \rightarrow \infty$, 
~(\ref{setone}) imposes $\alpha = \phi$. 
Substituting~(\ref{setone}) in~(\ref{settwo})
with $\alpha=\phi$, 
and redefining $\tilde\phi = 2 \phi$,  gives
\begin{equation}
\label{minone}
2 \Sigma\, m_u\, \sin(\tilde\phi) = 2 {\tau} \,(\theta - \tilde\phi).
\end{equation}
This minimizes
\begin{equation}
\label{nfone}
E(\theta) = - m_u \Sigma \cos\tilde\phi + {\tau\over 2} (\theta - \tilde \phi)^2,
\end{equation}
which is the potential energy for $N_f=1$, as expected. 
These limits,  $m_d \rightarrow 0$ and $m_d \rightarrow \infty$, 
illustrate that the dependence in $\theta$  ({\em i.e.} CP violation)
 is
controlled by the lightest quark flavour. 

\subsection{One  and zero flavour.}
We begin with the analysis 
of the one flavour case~(\ref{nfone}), which is analogous to the (bosonized) massive
Schwinger model in two dimensions~\cite{Coleman:1976uz}. At $\theta=0$, (\ref{minone}) 
gives (dropping the
tilde) 
\begin{equation} 
\sin\phi = - {\tau\over m_u \Sigma} \, \phi
\end{equation}
which is trivially satisfied for $\phi=0$. This solution is the true ground state
and is  CP conserving, in agreement with the Vafa-Witten theorem~\cite{Vafa:1984xg}. 
Other solutions are
possible if
\begin{equation}
\label{condone}
\tau/m\Sigma \lsim 2/3\pi.
\end{equation}
 These
are CP violating, but metastable (see figure 1). At fixed 
$m_u$, the condition~(\ref{condone}) can be satisfied in the very large $N_c$ limit, because
$\Sigma = {\cal
O}(N_c)$, if $m_u$ increases or if $\tau = {\cal O}(N_c^0)$ diminishes (dramatically) with respect 
to its phenomenological value.
 The first
case has been discussed by Witten~\cite{Witten:1980sp}. The latter possibility 
has been raised in the context of the
deconfining phase transition in QCD at finite temperature~\cite{Kharzeev:1998kz}. 
Finally, changing the mass quark allows to change the number of light flavours, as  exemplified in the previous section.

At $\theta=\pi$, the trivial, CP conserving solution is $\phi=\pi$. 
Spontaneous CP violation
can occur only if  %
\begin{equation}
\label{conpione}
\tau \lsim m_u \Sigma,
\end{equation}
in which case the CP conserving solution becomes a maximum (see
figure 1). The inequality (\ref{conpione}) is the equivalent of~(\ref{witten}) for $N_f=1$.
At $\theta = 2 \pi$, things are the same  as at $\theta=0$, but with $\phi$ shifted by $2 \pi$. Thus,
despite the presence of the term quadratic in $\theta$ in the potential energy~(\ref{nfone}), 
the ground state energy  is $2\pi$ periodic, simply because a shift like $\theta \rightarrow \theta +
 k\, 2 \pi$ 
 is reabsorbed in 
$\phi \rightarrow \phi + k \,2 \pi$. 

It may actually be worthwhile to emphasise 
that the Di Vecchia-Veneziano-Witten effective lagrangian is 
consistent  not only  with the  $\theta$ dependence  of QCD but {\em also}  of pure Yang-Mills theory.   
If we formally increase $m_u$, the system indeed shares 
some ressemblance with the limit of zero quark flavour. In particular, this limit 
provides a very simple derivation
 of  the peculiar $\theta$ dependence  of  large $N_c$ 
pure Yang-Mills theory. 
  Recently, Witten~\cite{Witten:1998uk} has derived the 
vacuum energy 
using the correspondence 
between large $N_c$ Yang-Mills theory and string theory on a 
certain space-time background~\cite{Maldacena:1997re}. (See also Shifman~\cite{Shifman:1998if} and  
Gabadadze~\cite{Gabadadze:1999na} for a discussion  in a field theory context.) 
The energy has a multibranch structure  
\begin{equation}
\label{branched}
E(\theta) = N_c^2 \,\mbox{min}_k \, F\left({\theta + k 2 \pi\over N_c}\right)
\end{equation}
where  $k$ is an integer and 
\begin{equation}
F\left({\theta \over N_c}\right) = C_0 + {\tau\over 2} 
{\theta^2\over N_c^2} + {\cal O}\left({\theta^4\over N_c^4}\right)
\end{equation}
This form of the energy has been postulated by Witten many 
years ago~\cite{Witten:1979vv} in order to reconcile  large $N_c$ 
 with the requisite  $2\pi$ periodicity in $\theta$. A striking feature 
of~(\ref{branched}) is that, at fixed $\theta$,
it implies the existence of  many 
 non-degenerate vacua 
in pure Yang-Mills theory at large $N_c$. It has been 
argued in \cite{Witten:1980sp}, and explicitely shown in the more 
recent \cite{Witten:1998uk} and \cite{Shifman:1998if}, that these states are 
stable at infinite $N_c$.  For completeness,
 we show how  this conclusion, as well as 
 the large $N_c$ scaling of the quantities of interest, can be reproduced 
within the framework of the Di Vecchia-Veneziano-Witten effective Lagrangian.

For $m_u \Sigma \gg \tau$, the potential has a large  number, $k_{\mbox{max}} \sim m\Sigma/\tau \sim N_c$, of local 
minima. At $\theta = 0$, the ground state is unique and CP conserving, $\phi_0 =0$. Then there
are two adjacent, degenerate, metastable solutions 
\begin{equation}
\phi^\pm_1 \approx  \pm 2 \pi ( 1  - \tau/m \Sigma) 
\end{equation}
with energy
\be
\Delta E_1 \approx 2  \pi^2 \tau
\ee
where we have substracted the trivial contribution from the quark mass term. For generic $\theta$ and 
$k \ll k_{\mbox{max}}$, $\phi_\pm \approx \pm 2 k \pi$, with $\Delta E_k \approx {\tau\over 2} (\theta - k\,  2\pi)^2$.
As $\theta$ increases from $0$, $\phi_0$ and the $\phi_k^-$ go up and the $\phi_k^+$ go down. At 
$\theta = \pi$, $\phi_0 \approx \pi \tau/m \Sigma$ and 
$\phi_1 \approx 2\pi - \pi \tau/m \Sigma$ become degenerate. In this picture, 
CP is spontaneously broken at $\theta = \pi$, but only very 
slightly as $\bar \theta \approx  \pm 2\pi \tau/m \Sigma$ modulo $\pi$, which goes to zero as $m$ or $N_c$ go to infinity.
 The height of the potential barrier between the two vacua is $m\Sigma \sim N_c$ and they can 
be related by a domain wall.~\footnote{In the AdS/CFT approach, the domain wall is a wrapped six-brane (with $\sigma \sim N_c$). 
In the $N=1$ SUSY approach, the domain wall  relates two adjacent  vacua,  out of the $Z_{N_c}$ distinct ones with different 
gluino condensate. } The  profile of the domain wall  
can be easely estimated in the limit $\tau \ll m \Sigma$.
Choosing  $\phi(x) = \phi_0 = 0 + {\cal O}(\tau/m \Sigma)$ at spatial
 $x= -\infty$ and $\phi(+\infty) = \phi_1 =  2 \pi + {\cal O}(\tau/m \Sigma)$ and 
solving the differential 
equation for the phase $\phi(x)$, 
\begin{equation}
f_\pi^2 \partial_x^2 \phi - {\delta_\phi E(\phi)} =0
\end{equation}
where $E$ is given by (\ref{nfone}), we find
\begin{equation}
\label{profile}
\phi(x) \approx \pi + 2 \arctan \left\{ {\exp{(\sqrt{m\Sigma} x /f_\pi}) - 1\over  \exp{(\sqrt{m\Sigma}\,x/f_\pi}) + 1}\right\}
\end{equation}
In the same approximation, the  tension of this domain wall is
\begin{equation}
\label{surftens}
\sigma = \int_{-\infty}^{+\infty} dx \, \left\{{1\over 2} f_\pi^2 (\partial_x \phi)^2 + E(\phi)\right\}
\approx 8 f_\pi \, \sqrt{m \Sigma} \sim N_c.
\end{equation}
Let us comment about the  
validity of these expressions,  (\ref{profile}) and  (\ref{surftens}).
Note that the domain wall configuration arises from the balance between the kinetic and quark mass terms, which are of 
the same order in the low energy expansion of the effective theory.\footnote{The situation is thus better than for skyrmions. The stability of these topological defects necessitates to mix different orders in the low energy expansion, while in 
the case of domain walls the calculations can be made arbitrarily reliable, even in an effective theory.} 
From (\ref{profile}),
the domain wall has a width $\delta \approx f_\pi/\sqrt{m \Sigma} \sim 1/M$, 
where $M \sim N_c^0$ is the meson mass, so that the gradient of the meson field is typically ${\cal O}(M/f_\pi)$ accross the wall. 
Corrections to the leading order results (\ref{profile}) and  (\ref{surftens}) from higher order operators 
are thus under control as long as
we keep $M \ll f_\pi$, which imposes the following hierarchy $\tau \ll m \Sigma \equiv M^2 f_\pi^2 \ll f_\pi^4$. 

\bigskip 

Increasing  $\theta$ further,  $\phi \approx 0$  
becomes a local minimum  while $\phi \approx 2 \pi$ becomes the true 
ground state. When $\theta$ reaches $2\pi$, 
things just repeat,   with $\phi$ shifted by $2 \pi$. If we integrate out the heavy quark, 
 the ground state energy as  function of $\theta$  becomes
\begin{eqnarray}
E(\theta) &=& {\tau\over 2}\,\mbox{min}_{\phi} \,(\theta - \phi)^2\nonumber\\
&\approx& {\tau\over 2}\,\mbox{min}_{k} \,(\theta - 2 k \pi)^2\nonumber
\end{eqnarray}
which agrees with the multibranch structure of~(\ref{branched}) and 
shows that the Di-Vecchia-Veneziano-Witten
is (at least formally) consistent with the
 expected $\theta$ dependence of pure Yang-Mills theory at large $N_c$.
~\footnote{For   
{\em this}  particular matter, the modifications of the Di Vecchia-Veneziano-Witten 
effective lagrangian proposed in~\cite{Fugleberg:1998kk} thus seem superfluous.}

At fixed $\theta$ there is 
a large number of metastable states, corresponding to $\phi \approx k 2 \pi$.~\footnote{In this
 picture, $\langle\phi\rangle$ is a sort of 
 auxiliary field which labels the different vacua branches. 
It plays a role similar to the gluino 
condensate in $N=1$ SUSY or  to the flux of the $U(1)$ gauge field from the Ramond-Ramond sector of type 
IIA superstring in the AdS/CFT correspondence.} (See figures 1 and 2.)
For $k$ large, $k\lsim k_{\mbox{max}}$, these states are essentially 
unstable.
For small $k \ll  k_{\mbox{max}}$, the 
lifetime of the lowest lying  solutions can be
 easely evaluated.  The energy difference between 
 two  adjacent ($\Delta k=1$) 
 vacua $\Delta E \approx 2 \tau \pi^2 \sim N_c^0$
 is  much less than the eight of the potential barrier, $E \approx m \Sigma \sim N_c$ and  the thin-wall
 approximation applies~\cite{Kobzarev:1975cp,Coleman:1977py}. The decay rate is
$\Gamma \exp(-S_E)$, where $S_E$ is the euclidean action for a bubble of lower energy vacuum. 
In the thin wall approximation, $S_E$ is well approximated by
\begin{equation}
\label{instanton}
S_E \approx -  \pi^4\, \tau\,{R^4\over 4} + 2 \pi^2 R^3\, \sigma,
\end{equation}
where the first term is the contribution from the volume 
 and the other from the surface tension~(\ref{surftens}), while 
$R$ is the radius of the 
bubble of ``true'' vacuum.
The action $S_E$~(\ref{instanton}) is extremized for 
\begin{equation}
R_c \approx {6 \sigma\over \pi^2 \tau} \sim N_c.
\end{equation}
while from~(\ref{profile}), the  bubble 
wall tickness  is $\delta \sim f_\pi/\sqrt{m\Sigma}  \sim N_c^0 \ll R_c$.
The rate for a (low lying) false vacuum to decay to its lower energy neighbour is finally given by
\begin{equation}
\label{rate}
\Gamma \sim \exp\left(- {27 \, 2^8\, f_\pi^4 \, m^2\,  \Sigma^2\over \pi^4 \tau^3}\right) \sim \exp(-N_c^4)
\end{equation}
so that, as $N_c$ goes to infinity, the non-degenerate vacua become stable. 

Shifman has also attempted to compute the decay rate of the lowest lying metastable states in the pure Yang-Mills theory, 
but starting from N=1 SYM and decoupling the 
gluinos~\cite{Shifman:1998if}. (See also Gabadadze~\cite{Gabadadze:1999na}.) His
expression, which  rests on the (well motivated) assumption that the domain walls relating adjacent vacua
in N=1 SYM are BPS saturated states, is reliable for small gluinos masses, {\em i.e.} $m_g$ smaller than $\vert 
\langle \lambda \lambda \rangle\vert = \Lambda$ where $\langle \lambda \lambda \rangle$ is the gluino condensate.
Similarly, 
the expression we have derived, (\ref{rate}), is valid as long as chiral perturbation theory is reliable, that is 
 if the quark mass is such that
$M^2 \ll f_\pi^2$, where $M$ is the meson mass. 
In order to match to the pure glue theory requires to 
decouple respectively  the gluinos or the heavy quark. In the former case, one looses the control of holomorphy, while in the
latter case chiral perturbation theory breaks down. As could be expected, 
the predictions of N=1 SYM and of chiral perturbation theory can only be compared at a qualitative level.
It is quite remarkable that the $N_c$ scaling of the decay rate,~(\ref{rate}), is precisely
 the same in both approaches.

\subsection{Two flavours, degenerate case.}
If $m_u=m_d\equiv m$, the equations~(\ref{setone}--\ref{settwo}) can be easely 
solved at $\theta=\pi$, at
least in the limit $m \Sigma \ll \tau$, which we will assume to hold from now on. 
Defining $\phi = \pi/2
+ \varphi$, the potential energy~(\ref{energy})
\begin{equation}
\label{Ephi}
E(\theta) = - 2\, m\, \Sigma\, \cos\phi \cos\alpha +  {\tau\over 2}\, (\theta - 2\phi)^2
\end{equation}
becomes
\begin{equation}
\label{epi}
E(\pi) =   2\, m\, \Sigma\, \sin\varphi\,\cos\alpha + 2\, \tau\, \varphi^2
\end{equation}
and~(\ref{setone}--\ref{settwo}),
\begin{eqnarray}
\label{setm}
\sin\varphi \, \sin\alpha &=& 0\\
\label{setmtwo}
\cos\varphi\,\cos\alpha &=& -2 {\tau\over m\Sigma} \varphi 
\end{eqnarray}

Note first that if $\tau/m\Sigma \rightarrow \infty$, $\varphi=0$, and  CP is not broken
($\bar \theta \equiv 2 \varphi =0$). The potential~(\ref{epi})  vanishes for any 
$\alpha$---the pion are massless, 
and the transition at $\theta=\pi$ 
is of second order. This is precisely the situation that occurs to 
leading order in $SU(2)\times SU(2)$ $\chi PT$ and raised as a puzzle by Smilga~\cite{Smilga:1998dh}. 
Integrating out 
the $\eta^\prime$ from~(\ref{Ephi}) in the limit $\tau/m\Sigma \rightarrow \infty$, sets
$\phi\equiv \theta/2$ and gives
\begin{equation}
\label{energycpt}
E(\theta) = - 2\, m\, \Sigma\, \cos{\theta\over 2}\, \cos\alpha.
\end{equation}
which is the potential energy to leading order in $\chi PT$. At $\theta=\pi$,~(\ref{energycpt})
vanishes.

\bigskip

In the limit $m \Sigma \ll \tau$, it is easy to solve the equations~(\ref{setone}). The 
trivial CP conserving
 solution is
\begin{equation}
\varphi_I=0 \quad , \quad \alpha_I= \pi/2\,\,  \mbox{modulo}\,\, \pi
\end{equation}
This solution is however a saddle point, with $E_I =0$. The true ground
state is CP violating, with the two solutions
\begin{equation}
\begin{array}{ccrcc}
\varphi_{II} & \approx & - {m\Sigma/ 2 \tau} & , &  \alpha_{II}=0\\
\varphi_{III} & \approx &  {m\Sigma/ 2 \tau} & , &  \alpha_{III}=\pi\\
\end{array}
\end{equation}
and
\begin{equation}
E_{II} = E_{III} \approx - {m^2\Sigma^2\over 2 \tau} 
\end{equation}
Contrary to the one flavour case~(\ref{conpione}), for $N_f=2$ spontaneous CP
violation occurs for any {\em finite} $\tau/m\Sigma$.  

The two CP violating vacua are separated by an energy barrier (see figure 3 with $z=1$) and can 
be related by a domain wall. Note that for $m_u=m_d$ the vacuum has an $SU(2)/U(1) \sim S^2$ degeneracy 
and, correspondingly,
there is an infinite family of domain walls with the same tension. 
Two of these vacua are shown in fig. 3, $z=1$. The vacua at $\alpha = 0$ and $\alpha = 2 \pi$ as 
well as the saddle points at $\alpha = \pi/2$ and $\alpha = 3 \pi/2$ 
are equivalent and can mapped 
onto each other by a Weyl reflection, $\alpha \rightarrow - \alpha$ (modulo $2 \pi$). This degeneracy is lifted for
any $m_u \not= m_d$: there are only {\em two} vacua, which are shown in figure 3 for $z\not= 1$, and 
{\em one} domain wall\cite{Smilga:1998dh}. As discussed in the next section,
there is also a wall-some sphaleron, {\em i.e.} a metastable configuration 
 interpolating between the two vacua, which relaxes into the domain wall if subject to perturbations. 
\footnote{We thank A. Smilga for making this point clear to us. Wall-some sphalerons have been 
first discussed in the context of supersymmetric effective theories.\cite{Smilga:1998vs} }

In the limit $m\Sigma \ll \tau$, $\varphi_{II,III} \ll 1$, and the domain wall profile 
is essentially along the
$\alpha$ direction. To estimate the tension of the domain wall, we first 
integrate out $\varphi$ using~(\ref{setmtwo})  
\begin{equation}
\label{pa}
\varphi \approx - {m \Sigma\over 2 \tau}\,\cos\alpha
\end{equation}
and substituting in~(\ref{epi}) to get the potential energy for $\alpha$:
\begin{equation}
\label{pot}
E(\alpha) = - {m^2\Sigma^2\over 2 \tau}\, \cos^2\alpha
\end{equation}
We can then compute the 
tension of the domain wall as in the previous section, to get
\begin{eqnarray}
\label{tension}
\sigma_\alpha &\approx&  {2 \,m\Sigma \,f_\pi\over \sqrt \tau} 
\end{eqnarray}
The error made by neglecting the gradient of the $\eta^\prime$ condensate $\varphi$ within the domain wall can be estimated using~(\ref{pa})
 to eliminate $\alpha$ instead. This gives
\begin{eqnarray}
\sigma_\varphi &\approx& 
 {\pi \,m^2\Sigma^2 f_\pi\over 4 \tau^{3/2}} 
\end{eqnarray}
which confirms that $\sigma_\varphi \ll \sigma_\alpha$ for $m\Sigma \ll \tau$. 

\bigskip

How do these results compare to the predictions of $\chi PT$ ?
At leading order the potential vanishes at $\theta=\pi$. As discussed above, this corresponds to
the  limit $\tau \rightarrow \infty$ or $m_{\eta^\prime} \rightarrow \infty$.
Obviously, there is no trace of an $\eta^\prime$ condensate. 
As shown by Smilga, at next-to-leading order the only ${\cal O}(p^4)$ operator relevant at
$\theta=\pi$ is $O_7$  (following the nomenclature of Gasser and
Leutwyler~\cite{Gasser:1984yg}). Adding $O_7$ to (\ref{energycpt}) gives
\begin{equation}
E(\alpha) = - 2 m \, \Sigma\, \cos{\theta\over 2}\, \cos\alpha -
 4 l_7 \left({m\Sigma\over f_\pi^2}\right)^2 \,
\sin^2{\theta\over 2}\,\cos^2\alpha 
\end{equation}
or, at $\theta=\pi$,
\begin{equation}
\label{ecptnlo}
E(\alpha) = - 4 l_7 \left({m\Sigma\over f_\pi^2}\right)^2\,\cos^2\alpha 
\end{equation}
Furthermore, in the large $N_c$ limit, the $\eta^\prime$ is 
 ``not that heavy''
$m_{\eta^\prime}^2 \sim 1/N_c$, and the coupling $l_7$ can be saturated by $\eta^\prime$ meson
exchange~\cite{Gasser:1984yg}, 
\begin{equation}
\label{lseven}
l_7 = {f_\pi^2\over 2 m_{\eta^\prime}^2} = {f_\pi^4\over 8 \tau}
\end{equation}
where we used  the Veneziano-Witten relation~(\ref{vw}).\footnote{In $SU(2)\times
SU(2)$ chiral perturbation theory at finite $N_c$ and finite strange quark mass, 
$l_7$ is instead saturated by  $\eta$ meson
exchange, $l_7\propto 1/m_\eta^2\sim 1/m_s$. Although we have not pursued in this direction, most 
presumably
the results of next-to-leading order  $SU(2)\times
SU(2)$  chiral perturbation
theory could be recovered starting from the $SU(3)\times SU(3)$ case to leading order, and decoupling the strange
quark. In the latter case, the vacuum degeneracy is indeed lifted already at leading order in
$\chi PT$. Incidentally, this is precisely the reason why $\tau$ does not appear in the inequality~(\ref{witten}).
 See the discussion of Smilga.} Substituting~(\ref{lseven}) in~(\ref{ecptnlo}) gives back
the large $N_c$  prediction~(\ref{pot}). Also,~(\ref{tension}) reads
\begin{equation}
\sigma_\alpha \approx {m\Sigma\over f_\pi}\sqrt{32 l_7}
\end{equation}
which is precisely the result derived by Smilga.

\bigskip

That predictions from next-to-leading order chiral $\chi PT$ and 
the large $N_c$ effective lagrangian can be made to agree is a nice consistency check. As a
 bonus, in the large $N_c$ framework, we see explicitely how the two 
degenerate vacua are related to spontaneous CP breaking with an $\eta^\prime$ condensate, and why they disappear as
 $\tau/m\Sigma\rightarrow \infty$, with a second order transition at $\theta =\pi$,
features which, for obvious reasons,
 are not manifest within $\chi PT$.~\footnote{As emphasized in~\cite{Peris:1995dh}, $l_7$ is
anomalously large in the large $N_c$ framework, $l_7 ={\cal O}( N_c^2)$. If large $N_c$ is adopted as
a guideline, consistency would require to work with the extended  symmetry $U(N_f)\times U(N_f)$, {\em
i.e.} with a dynamical $\eta^\prime$, rather than $SU(N_f)\times SU(N_f)$. Our discussion provides
 another illustration of this point.}

\subsection{Two flavours, mass splitting effects.}

If $m_u\not=m_d$, the algebra is just a bit more cumbersome. Defining 
\begin{equation}
z=m_d/m_u\, ,
\end{equation}
\begin{equation}
y=2\tau/\Sigma m_u
\end{equation}
and $\phi = \varphi + \pi/2$, the potential energy~(\ref{energy}) at
$\theta =\pi$ becomes
\begin{equation}
\label{tildeE}
E = \Sigma m_u \left\{ (1+z)\sin\varphi\cos\alpha + (1-z)\cos\varphi\sin\alpha + y\,
\varphi^2 \right\}
\end{equation}
In what follows, we will keep $y$ fixed and vary $z$ ({\em i.e.} $m_d$).
Minimizing with respect to $\varphi$ and $\alpha$ gives
\begin{eqnarray}
\label{cosalphaz}
\sin\alpha &=& - y\,{1-z\over 2 z} \,{\varphi\over \sin\varphi}
\end{eqnarray}
while eliminating $\alpha$, gives 
\begin{equation}
{\sin^2\varphi\over \varphi^2} \,{4 z^2\over y^2} = (1-z)^2 + (1+ z)^2 \tan^2\varphi
\end{equation}
As for  $z=1$, the CP conserving solution is again 
\begin{equation}
\varphi_I = 0 \quad \mbox \quad \alpha_I = \pi/2 \,\, \mbox{modulo}\,\,\pi
\end{equation}
Note however that, for $z\not=1$, these are two distinct solutions  with   energies
\begin{eqnarray}
\label{lowestE}
E_{Ia} &=& (1-z) m_u \Sigma \equiv (m_u - m_d) \Sigma \nonumber\\
E_{Ib} &=& - (1-z) m_u \Sigma \equiv (m_d -m_u) \Sigma
\end{eqnarray}
For $m_d>m_u$ (respectively $m_d <m_u$), $E_{Ia}$ ($E_{Ib}$) has lower energy. 

In the limit $m\Sigma \ll \tau$, we can write down the two CP violating solutions
\begin{equation}
\label{phiz}
\vert \varphi\vert \approx {1\over 1+z} \sqrt{4 z^2/y^2 - (1-z)^2}
\end{equation}
These exist provided 
\begin{equation}
\left({2 z^\ast\over y}\right)^2 < (1-z_\ast)^2,
\end{equation}
which gives two critical values of the mass ratio $z=m_d/m_u$, 
\begin{eqnarray}
z_1^\ast &=&{y\over y +2}\,< 1 \\
\nonumber \\
z_2^\ast &=& {y\over y-2} \,> \,1 
\end{eqnarray}
For two light flavours, CP is spontaneously broken with a first order 
phase transition at $\theta=\pi$  if and only if~\footnote{That the two solutions $z_1^\ast$ and
$z_2^\ast$ are not  symmetric around $z=1$ is not surprising. At {\em fixed} $m_u$, the two limits,
$z\rightarrow \infty$ and $z\rightarrow 0$ correspond to two physically different situations:
decoupling of a heavy quark in the  former case ($N_f \rightarrow 1$) and $N_f=2$ with a massless
quark in the latter. The distinction goes however away for larger $y$'s, $z^\ast \approx 1 \pm 2/y$.
If we exchange the role of $m_d$ and $m_u$,  
$z \rightarrow z^{-1}$, $y \rightarrow z y$, and $z_1^\ast \leftrightarrow z_2^\ast$.
}
\begin{equation}
\label{tytgat}
z_1^\ast < {m_d\over m_u} < z_2^\ast
\end{equation}
which is the equivalent for two flavours of Witten's
inequality~(\ref{witten}). 
For realistic quark mass and $\tau$,~(\ref{tytgat})
gives
\begin{equation}
\left\vert{ m_d -m_u\over m_u + m_d}\right\vert \lsim  {m_\pi^2\over
m_{\eta^\prime}^2}
\end{equation}
which, just like~(\ref{witten}), is unfortunately not satisfied in Nature.~\footnote{Using the mass of $\pi^0$ 
and $\eta^\prime$, $\tau \sim (200 MeV)^4$ and $\Sigma \sim (250 MeV)^3$ (for $N_f= 3$).  As 
$m_u \approx 4 MeV$ and $m_d \approx 7 MeV$, $y\sim 40$ and $z^\ast_{1,2} \sim 1 \pm 1/20$, to be compared to $m_d/m_u \approx 1.75$. }

\bigskip 

For completeness, we give  the CP violating solutions for
 $z$ in the range of~(\ref{tytgat}). Assuming small mass splitting, so that
$\alpha \ll 1$ modulo
$\pi$, and using~(\ref{phiz}) and~(\ref{cosalphaz}), we have
\begin{eqnarray}
\varphi_{II} &\approx& -\vert \varphi\vert \quad \mbox{and}\quad \alpha_{II} \approx -{y\over 2
z}\,(1-z) \nonumber\\
\varphi_{III} &\approx& \vert \varphi \vert 
\quad \mbox{and}\quad  \alpha_{III} \approx \pi + {y\over
2 z}\,(1-z) 
\end{eqnarray}
For larger mass splittings, CP violation goes away ($\varphi \rightarrow 0$) and
\begin{equation}
\label{migration}
\alpha_{II,III} \rightarrow    -{\pi\over 2} \,\,\mbox{mod}\,\, \pi \quad \mbox{if}\quad  z\rightarrow z_1^\ast
\end{equation}
or
\begin{equation}
\alpha_{II,III} \rightarrow  {\pi\over 2} \,\,\mbox{mod}\,\, \pi \quad \mbox{if}\quad z\rightarrow z_2^\ast\nonumber
\end{equation}
The energy difference $\Delta E$ between the CP violating vacua and the {\em lowest energy} CP conserving saddle point
of~(\ref{lowestE}) is
\begin{equation}
\Delta E_{II}= \Delta E_{III} \approx - {\Sigma^2 m_u^2\over 2 \tau} \,\left({z^\ast - z\over z^\ast
- 1}\right)^2
\end{equation}
which  vanishes at $z =z_1^\ast$ or $ z = z_2^\ast$. 
For $z \approx 1$, the tension of the domain wall relating the two CP violating vacua is well
approximated by  
\begin{equation}
\sigma \approx {2 m_u \Sigma f_\pi\over\sqrt{\tau}}\,\left\vert{ z^\ast - z\over z^\ast - 1}\right\vert
\end{equation}
where, again, we are neglecting a small contribution from  the $\eta^\prime$ condensate. (Compare with~(\ref{tension}).) As $z$ approaches $z_{1,2}^\ast$, the CP violating vacua merge
into a unique, CP conserving vacuum and the domain wall disappears: the phase transition is of second order, 
with massless pion excitations. In particular,  chiral perturbation theory at leading order corresponds to the limit $\tau \rightarrow \infty$, or  $z_1^\ast = z_2^\ast = 1$.

As can be seen on figure 3, for $z\not= 1$ there are apparently two distinct potential barriers between the two CP violating vacua, which correspond to the two CP conserving saddle points of~(\ref{lowestE}). The  
configuration that relates the two vacua  by going  through the saddle point of higher energy 
at $\phi=0$ and $\alpha = \pi/2$ modulo $\pi$~(\ref{lowestE})  is not a domain wall but  a wall-some sphaleron, {\em i.e.}
a metastable configuration which, if subject to the slightest perturbation, will relax to the domain wall configuration. 
This sphaleron is a remnant of the 
infinite family of equivalent domain walls that exist in the degenerate limit $m_u=m_d$. Similar objects have been 
encountered in supersymmetric theories~\cite{Smilga:1998cx,Smilga:1998vs}.

\section{Summary: phase diagram of  QCD in the $z-\theta$ plane.} 
\label{conclusion} %

For $N_f=2$, we have  shown that CP is spontaneously broken at $\theta =\pi$ also for  finite quark mass
splitting, and derived the inequality~(\ref{tytgat}), valid in the 
limit $m\Sigma \ll \tau$ ($m_\pi^2 \ll m_{\eta_\prime}^2$).
The resulting phase
diagram of $N_f= 2$ QCD in the $z-\theta$ plane, where $m_u$ and $\tau$ are held
fixed and $m_d$ is allowed to vary, is shown on figure 4. 
At $\theta = \pi$ there is a line of first order transitions for $z_1^\ast < m_d/m_u < z_2^\ast$. 
At the critical points $z^\ast$, the phase transition becomes of
second order, while beyond these, there is just a
smooth crossover.  
If the ratio $y = \tau/2 m_u \Sigma$ increases, the critical lines  shrinks to a
point at $z^\ast =1$, with a second order phase transition. This is the limit described
by leading order $\chi PT$.
In the opposite limit of $m_u\Sigma \gsim \tau$, like at very large $N_c$ or for heavy quarks, the critical line is
stretched.  If  $m_d$ decreases, spontaneous CP violation still goes away at some $0 <z_1^\ast < 1$. 
If  $m_d$ increases instead, the ``down quark'' eventually decouples
 and the system
is essentially that of one flavour, $N_f = 1$. In this case, if $m_u\Sigma \gsim \tau$, 
there is spontaneous CP
violation at $\theta = \pi$ for any $m_d$ and  $z_2^\ast$ goes to
infinity.
Finally,
if both quarks are heavy, the system is analogous to pure Yang-Mills theory, and CP is always broken at $\theta = \pi$.
This complete our survey of Dashen's phenomenon for $N_f=0,\,1,\,2$.  

\acknowledgements

We 
would like to thanks Andrei Smilga and Robert D. Pisarski for useful comments, 
Rafel Escribano and Fu-Sin Ling for discussions and
the FNRS for financial support.

\begin{figure}[h]
\label{f0}
\begin{center}
\epsfig{file=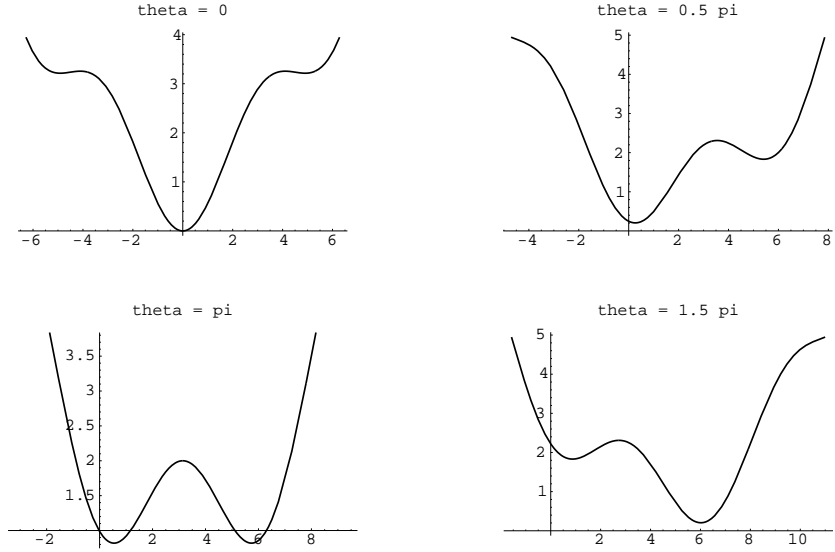,height=8cm}
\caption{Potential energy for $N_f=1$ as  function of the vev of  $\eta^\prime$, $\phi$. 
For $m\Sigma \gsim \tau$, there
are two metastable CP violating solutions at $\theta=0$. At $\theta=\pi$, there are two degenerate vacua CP violating vacua. 
At $\theta=2 \pi$, the potential is the same as at $\theta=0$, but shifted by $2 \pi$. (See figure 2.)}
\end{center}
\end{figure}  

\begin{figure}[h]
\label{fone}
\begin{center}
\epsfig{file=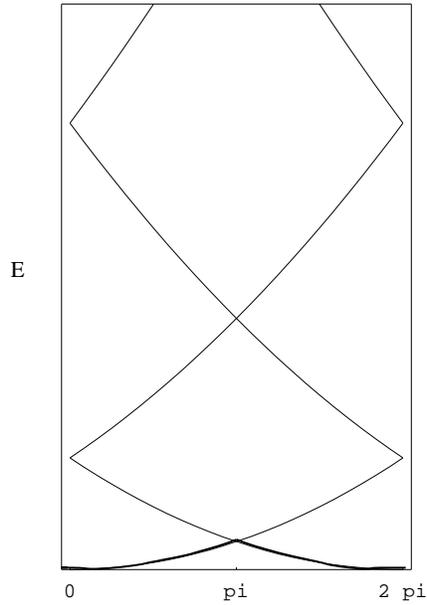,height=8cm}
\caption{Energy levels for $m\Sigma \gg \tau$ ($\sim N_f=0$) and $ 0 \leq \theta < 2 \pi$ (first Brillouin zone).
For fixed $\theta$, there are many {\em metastable} 
vacua, corresponding to $\langle \phi\rangle \approx k 2 \pi$. At $\theta=\pi$, the lowest energy levels cross and
 there are two degenerate  vacua. (See figure 1.) }
\end{center}
\end{figure}

\begin{figure}[h]
\label{figtwo}
\begin{center}
\epsfig{file=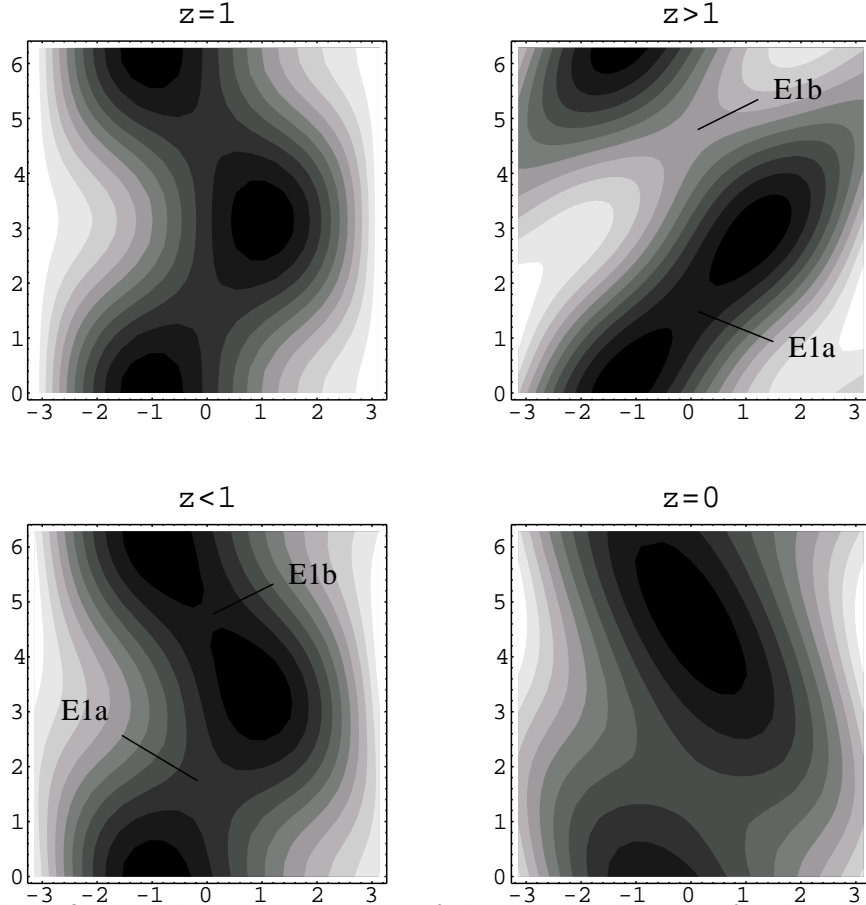,height=12cm}
\caption{These figures show contour plots of the potential energy for the case of two quark flavours at $\theta = \pi$. 
The vacuum expectation of the $\eta^\prime$ field, $\varphi = \phi - \pi/2$,
 is plotted on the horizontal axis. The CP violating vacua correspond to the dark spots with $\varphi \not= 0$. The vev of the  pion field, $\alpha$, is plotted on the vertical axis. 
(We have taken $m \Sigma \sim \tau$ to make the picture more impressive.) At $z=1$ ($m_u=m_d$)
 there are two CP violating vacua, 
separated by an energy barrier. Note that the vacua at $\alpha \sim 2 \pi$ and $\alpha \sim 0$ are {\em identical}. 
In this degenerate limit, there is an infinite family of equivalent vacua, related by an infinite number of equivalent domain wall configurations. 
At $z\not=0$, this degeneracy is lifted and there are only two vacua. Consider $z>1$ for definitiveness. The configuration which 
interpolates between the two vacua by going through the saddle with energy $E_{1a}$ (see~(\ref{lowestE})) is the domain wall.
 The one that goes through the saddle with energy $E_{1b} > E_{1a}$ is a wall-some sphaleron.
} 
\end{center}
\end{figure}

\begin{figure}
\label{figone}
\begin{center}
\epsfig{file=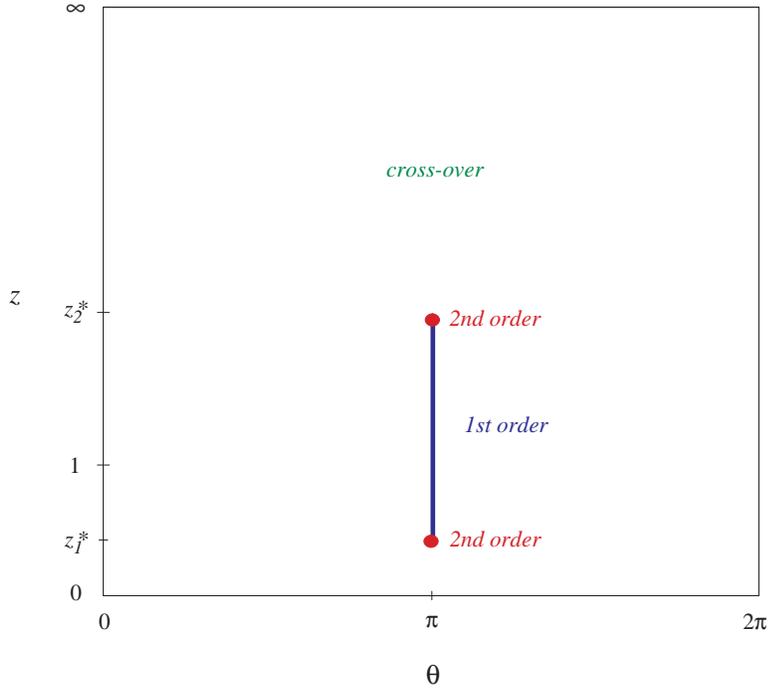,height=10cm}
\caption{Phase diagram of $N_f=2$ QCD in the $z-\theta$ plane, with $z = m_d/m_u$, for generic, {\em fixed} $m_u$
and $\tau$.  Spontaneous CP violation occurs on the line of first order 
phase transitions at $\theta=\pi$ for $z_1^\ast < z < z_2^\ast$. Chiral perturbation theory at leading order corresponds to
 the particular limit in which $\tau \rightarrow \infty$ and  $z_1^\ast = z_2^\ast = 1$. If the quark mass increase
 for fixed $\tau$, the critical line is stretched. In particular, $z_2^\ast \rightarrow \infty$ if $m_d \gg m_u$ ($N_f \sim 1$) and $m_u \gsim \tau/\Sigma$.} 
\end{center} 
\end{figure}

\end{document}